\documentclass[aps,twocolumn,showpacs,floatfix]{revtex4}
\usepackage{graphicx}

\begin{document}
\title{Scaling $p_T$ distributions for $p$ and $\bar{p}$
produced  in Au+Au collisions at RHIC}
\author{W.C. Zhang, Y. Zeng, W.X. Nie, L.L. Zhu and C.B. Yang}
\affiliation{Institute of Particle Physics, Hua-Zhong Normal University,
Wuhan 430079, P.R. China}

\begin{abstract}
With the experimental data from STAR and PHENIX on the centrality
dependence of the $p_T$ spectra of protons and anti-protons produced
at mid-rapidity in Au+Au collisions at 200 GeV, we show that
for protons and anti-protons there exists a scaling distribution independent of the colliding
centrality. The scaling functions can also describe data from BRAHMS for both proton
and anti-proton spectra at $y=2.2$ and 3.2. The scaling behaviors are shown to be
incompatible with the usual string fragmentation scenario for particle production.

\end{abstract}

\maketitle

\section{Introduction}
One of the most important quantities in investigating properties of
the medium produced in high energy collisions is the particle distribution
for different species of final state particles. RHIC experiments have found
a lot of novel phenomena from the particle spectra, such as the unexpectedly large
$p/\pi$ ratio at $p_T\sim 3\ {\rm GeV}/c$ \cite{p_pi}, the constituent quark number scaling
of the elliptic flows \cite{flow}, and strong nuclear suppression of the pion spectrum
in central Au+Au collisions \cite{supp}, etc. From the spectrum one can learn a lot on the
dynamics for particle production.

In many studies, searching for a scaling behavior of some quantities vs suitable variables
is useful for unveiling potential universal dynamics. A typical example
is the proposal of the parton model from the $x$-scaling of the structure
functions in deep-inelastic scatterings \cite{xscal}. Quite recently,
a scaling behavior \cite{hy1} of the pion spectrum at mid-rapidity in Au+Au collisions
at RHIC was found, which related spectra with different collision centralities.
In \cite{yang1} the scaling behavior was extended to non-central region, up to
$\eta=3.2$ for both Au+Au and d+Au collisions. The same scaling function can be
used to describe pion spectra for $p_T$ up to a few GeV$/c$ from different colliding
systems at different rapidities and centralities. The shape of pion spectrum in those
collisions is determined by only one parameter $\langle p_T\rangle$, the mean
transverse momentum of the particle. It is very interesting to ask whether similar
scaling behaviors can be found for spectra of other particles produced in Au+Au
collisions at RHIC. In this paper, the scaling property of the spectra for protons
and anti-protons is investigated and compared with that for pions.

The organization of this paper is as follows. In Sec. II we will address the procedures
for searching the scaling behaviors. Then in Sec. III the scaling properties of
the spectra for protons and anti-protons produced in Au+Au collisions at RHIC
at $\sqrt{s_{NN}}=200\ {\rm GeV}$ will be studied. We discuss mainly the centrality
scaling of the spectra at mid-rapidity and extend the discussion to very forward
region with rapidity $y=2.2$ and 3.2 briefly. Sec. IV
is for discussions on the relation between the scaling behaviors and the string fragmentation
scenario.

\section{Method for searching the scaling behavior of the spectrum}
As done in \cite{hy1,yang1}, the scaling behavior of a set of spectra at different
centralities can be searched in a few steps. First, we define a scaled variable
\begin{equation}
z=p_T/K\ ,
\end{equation}
and the scaled spectrum
\begin{equation}
\Phi(z)=A\left.\frac{d^2N}{2\pi p_Tdp_Tdy}\right|_{p_T=Kz}\ ,
\end{equation}
with $K$ and $A$ free parameters. As a convention, we choose $K=A=1$ for the most central
collisions. With this choice $\Phi(z)$ is nothing but the $p_T$ distribution for the most
central collisions. For the spectra with other centralities, we try to coalesce all data
points to one curve by choosing proper parameters $A$ and $K$. If this can be achieved,
a scaling behavior is found. The detailed expression of the scaling function depends, of course,
on the choice of $A$ and $K$ for the most central collisions. This arbitrary can be overcome
by introducing another scaling variable
\begin{equation}
u=z/\langle z\rangle=p_T/\langle p_T\rangle\ ,
\label{eq3}
\end{equation}
and the normalized scaling function
\begin{equation}
\Psi(u)=\langle z\rangle^2\Phi(\langle z\rangle u)/\int_0^\infty \Phi(z)zdz\ .
\end{equation}
Here $\langle z\rangle$ is defined as
\begin{equation}
\langle z\rangle\equiv \int_0^\infty z\Phi(z)zdz/\int_0^\infty \Phi(z)zdz \ .
\end{equation}
By definition, $\int_0^\infty \Psi(u) udu=\int_0^\infty u\Psi(u) udu=1$.
This scaled transverse momentum distribution is in essence similar to
the KNO-scaling \cite{kno} on multiplicity distribution.

\section{Scaling behaviors of proton and anti-proton distributions}
\begin{figure}[tbph]
\includegraphics[width=0.45\textwidth]{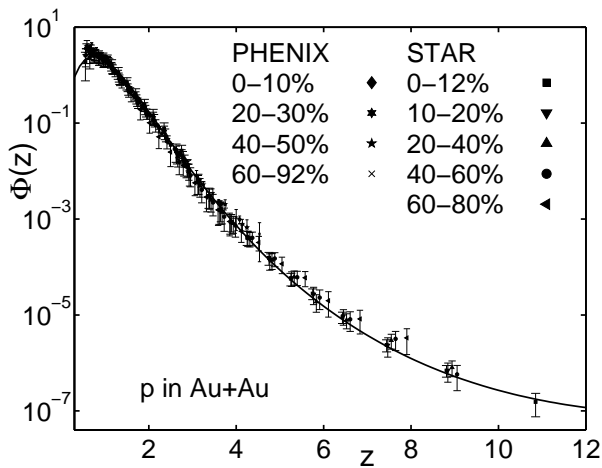}
\caption{Scaling behavior of the spectrum for protons produced at mid-rapidity in
Au+Au collisions at RHIC. The data are taken from \cite{exp,star}.
Feed-down corrections are considered in the data. The solid curve is from Eq. (6). }
\label{fig1}
\end{figure}

Now we focus on the spectra of protons and anti-protons produced at mid-rapidity
in Au+Au collisions at $\sqrt{s_{NN}}=200\ {\rm GeV}$. STAR and PHENIX Collaborations
at RHIC published spectra for protons and anti-protons at mid-rapidity
for a set of colliding centralities  \cite{exp,star}.
STAR data have a $p_T$ coverage larger than PHENIX ones.
As shown in Fig. \ref{fig1}, all data points for proton spectra at different centralities
can be put to the same curve with suitably chosen $A$ and $K$, by the procedure explained
in last section. The parameters are shown in Table I. Except a few points
for very peripheral collisions (centralities 60-92\% for PHENIX data and 60-80\%
for STAR data), all points agree well with the curve in about six orders of magnitude.
The larger deviation of data at centralities 60-92\% for PHENIX and 60-80\% for STAR
from the scaling curve may be
due to the larger centrality coverage, because the size of colliding system changes
dramatically in those centrality bins. For simplicity we define $v=\ln(1+z)$,
and the curve can be parameterized as
\begin{equation}
\Phi_p(z)=0.052\exp(14.9v-16.2v^2+3.3v^3)\ .
\end{equation}

\begin{figure}[tbph]
\includegraphics[width=0.45\textwidth]{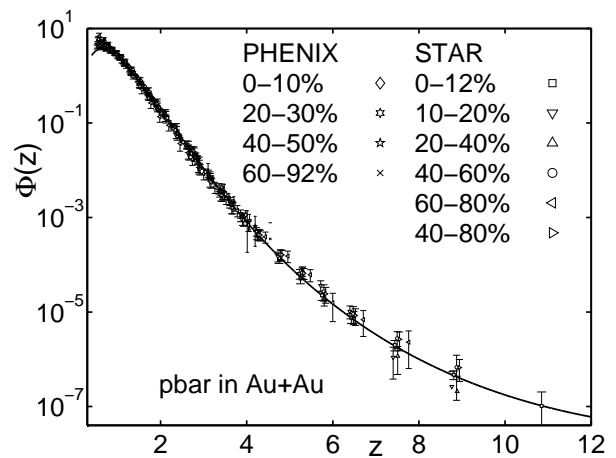}
\caption{Scaling behavior of the spectrum for anti-protons produced at mid-rapidity
in Au+Au collisions at RHIC. The data are taken from \cite{exp,star}.
Feed-down effects are not corrected in the STAR data for
$\bar{p}$. The solid curve is from Eq. (7).}
\label{fig2}
\end{figure}

Similarly, one can put all data points for anti-proton spectra at different centralities
to a curve with other sets of parameters $A$ and $K$ which are given also in TABLE I.
The agreement is good, as can be seen from Fig. \ref{fig2}, with only a few points in
small $p_T$ region for peripheral collisions departing a little from the curve.
For anti-proton the scaling function is
\begin{equation}
\Phi_{\bar p}(z)=0.16\exp(13v-14.9v^2+2.9v^3)\ ,
\end{equation}
with $v$ defined above.

\begin{table}
\def\tabcolsep{0.3cm}
\begin{tabular}{||c|c|c|c|c||}
\hline
STAR & \multicolumn{2}{c|}{$p$} & \multicolumn{2}{c||}{$\bar{p}$} \\ \hline
centrality & $K$ & $A$ & $K$ & $A$\\ \hline
0-12\% & 1 & 1 & 1 & 1\\ \hline
10-20\% & 0.997 & 1.203 & 1.005 & 1.417\\ \hline
20-40\% & 0.986 & 2.009 & 0.991 & 2.305\\ \hline
40-60\% & 0.973 & 4.432 & 0.993 & 5.414\\ \hline
60-80\% & 0.941 & 13.591 & 0.959 & 16.686 \\ \hline
40-80\% & & & 0.986 & 8.126 \\ \hline
 PHENIX & \multicolumn{2}{c|}{$p$} & \multicolumn{2}{c||}{$\bar{p}$} \\ \hline
 centrality & $K$ & $A$ & $K$ & $A$\\ \hline
 0-10\% & 1.042 &  1.226 & 1.068 & 2.404 \\ \hline
 20-30\% & 1.026 &  2.532 & 1.045 & 4.901\\ \hline
 40-50\% & 1.031 & 6.253 & 1.013 & 11.754 \\ \hline
60-92\% & 0.934 & 39.056 & 0.935 & 69.31 \\ \hline
BRAHMS & \multicolumn{2}{c|}{$p$} & \multicolumn{2}{c||}{$\bar{p}$} \\ \hline
 centrality & $K$ & $A$ & $K$ & $A$\\ \hline
 $y=2.2$ & & & 0.930 & 0.921\\ \hline
 $y=3.2$ & 1.079 & 0.754 & 1.153 & 6.985\\ \hline
\end{tabular}
\caption{Parameters for coalescing all data points to the same curves in Figs. 1 and 2.}
\end{table}
\begin{figure}[tbph]
\includegraphics[width=0.45\textwidth]{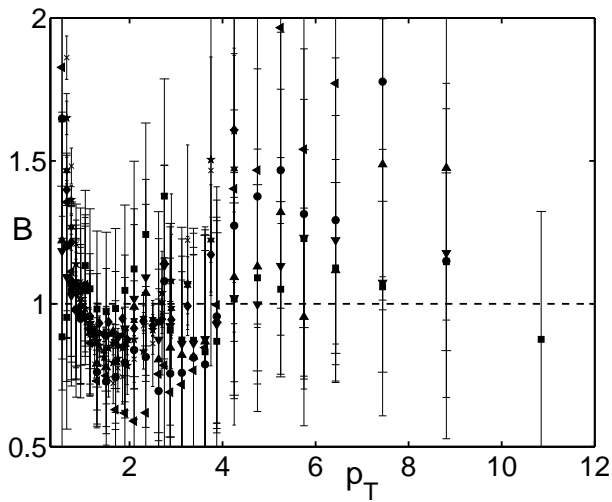}
\caption{Ratio between experimental data and the fitted results shown in Fig. 1.
STAR and PHENIX data are taken from \cite{exp,star}. Symbols are the same as
in Fig. 1.}
\label{fig3}
\end{figure}
To see how good is the agreement between the fitted curves in Figs. 1 and 2
and the experimental data, one can calculate a ratio
$$B={\rm experimental \ \ data}/{\rm fitted\ \ results}\ ,$$
and show $B$ as a function of $p_T$ in linear scale for all the data sets, as shown
in Fig. 3 for the case of proton. From the figure one can see that almost all the points
have values of $B$ within 0.7 to 1.3, which means that the scaling
is true within an accuracy of 30\%. This is quite a good fit, considering the fact
that the data cover about 6 orders of magnitude. For anti-protons, the agreement is
better than for protons.

\begin{figure}[tbph]
\includegraphics[width=0.45\textwidth]{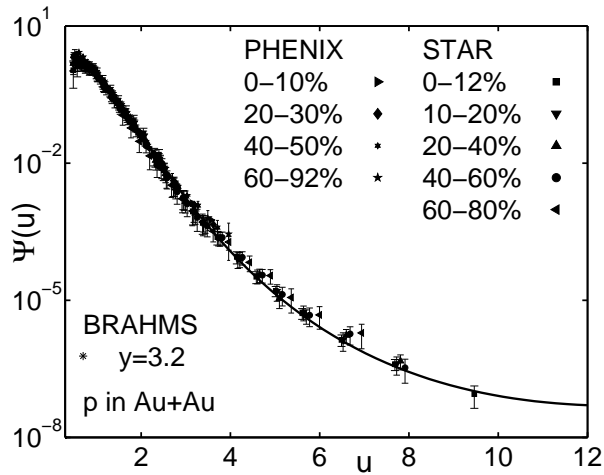}
\caption{Normalized scaling distribution for protons produced at mid-rapidity and very
forward direction in Au+Au collisions at RHIC with the scaling variable $u$.
STAR and PHENIX data are taken from \cite{exp,star} and BRAHMS data from \cite{brahms}.}
\label{fig4}
\end{figure}

Now one can see that the transverse momentum distributions for protons
and anti-protons satisfy a scaling law. For large $p_T$ (thus large $z$)
the scaling functions in Eqs. (6) and (7) behave as powers of $p_T$, though
the expressions are not in powers of $z$ or $p_T$.
The scaling functions in Eqs. (6) and (7) depend on the choices
of $A$ and $K$ for the case with centrality 0-12\% for STAR data.
With the variable $u$ defined in Eq. (\ref{eq3}) this dependence can
be circumvented. $\langle z\rangle$'s for protons and anti-protons are 1.14 and
1.08, respectively, with integration over $z$ in the range from 0 to 12, roughly
corresponding to the $p_T$ range measured by STAR. The normalized scaling functions
$\Psi(u)$ for protons and
anti-protons can be obtained easily from Eqs. (6) and (7) and are shown in
Figs. 4 and 5, respectively together with scaled data points as in Figs. 1 and 2.
A simple parameterization for the two normalized scaling  functions in Figs. 4
and 5 can be given as follows
\begin{eqnarray*}
\Psi_p(u)& = & 0.064\exp(13.6v-16.67v^2+3.6v^3)\ ,\\
\Psi_{\bar p}(u)& = & 0.086\exp(12.41v-15.31v^2+3.16v^3)\ ,
\end{eqnarray*}
with $v=\ln(1+u)$.

\begin{figure}[tbph]
\includegraphics[width=0.45\textwidth]{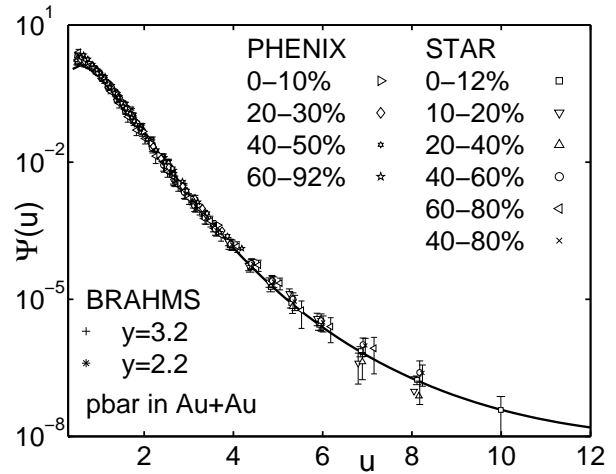}
\caption{Normalized scaling distribution for anti-protons produced at mid-rapidity and
very forward direction in Au+Au collisions at RHIC with the scaling variable $u$.
STAR and PHENIX data are taken from \cite{exp,star} and BRAHMS data from \cite{brahms}.}
\label{fig5}
\end{figure}

  As in the case for pion distributions, one can also investigate the $p_T$ distributions of
protons and anti-protons in non-central rapidity regions in Au+Au collisions. The only data
set we can find is from BRAHMS \cite{brahms} at rapidity $y=2.2$ and 3.2 with centrality
0-10\%. It is found that the BRAHMS data can also be put to the same scaling curves, as
shown in Figs. 4 and 5. The values of corresponding parameters $A$ and $K$ are also given
in TABLE I. Thus the scaling distributions found in this paper may be valid
in both central and very forward regions for protons and anti-protons produced in Au+Au
collisions at RHIC at $\sqrt{s_{NN}}$=200 GeV.

  Now one can ask for the difference between the scaling functions for protons and
anti-protons. After normalization to 1 the difference between the scaling distributions
$\Psi(u)$ for protons and anti-protons is shown in Fig. 6. In log scale the difference
between the two scaling functions is invisible at low $u$. To show the difference
clearly a ratio $r=\Psi_p(u)/\Psi_{\bar{p}}(u)$ is plotted in the inset of Fig. 6
as a function of $u$. The increase of $r$ with $u$ is in agreement qualitatively with data
shown in \cite{star} where it is shown that $\bar{p}/p$ decreases with $p_T$ monotonically.
The difference in the two scaling functions can be understood physically.
In Au+Au collisions there are much more quarks $u, d$ than
$\bar{u}$ and $\bar{d}$ in the initial state. In the central region in the
state just before hadronization, more $u$ and $d$ quarks can be found because
of the nuclear stopping effect in the interactions. As a consequence, more protons
can be formed from the almost thermalized quark medium than anti-protons in the small
$p_T$ regime. Experimental data show that in low $p_T$ region the yield of anti-proton
is about 80\% that of protons in central Au+Au collisions at RHIC. This difference
contributes to the net baryon density in the central region in Au+Au collisions at
RHIC. On the other hand, in the large $p_T$ region, protons and
anti-protons are formed mainly from fragmentation of hard partons produced in
the QCD interactions with large momentum transfer. As shown in \cite{fries},
the gluon yield from hard processes is about five times that of $u$ and $d$ quarks.
The fragmentation from a gluon to $p$ and $\bar{p}$ is the same.
The amount of $u, d$ quarks from hard processes is about 10 times that of $\bar{u},\bar{d}$
when the hard parton's transverse momentum is high enough.
It is well-known that the
fragmentation function for a gluon to $p$ or $\bar{p}$ is much smaller than that
for a $u$ or $d$ ($\bar{u}$ or $\bar{d}$) to $p$ ($\bar{p}$) because of the
dominant valence quark contribution to the latter process. As a result, the ratio of
yields of proton over anti-proton at large $p_T$ is even more than that at small $p_T$.
After normalizing the distributions to the scaling functions the yield ratio of
proton over anti-proton increases approximately linearly with $u$ when $u$ is large.
 It should be mentioned that no
such difference for $\pi^+, \pi^-$ and $\pi^0$, because they all are composed of
a quark and an antiquark.
\begin{figure}[tbph]
\includegraphics[width=0.45\textwidth]{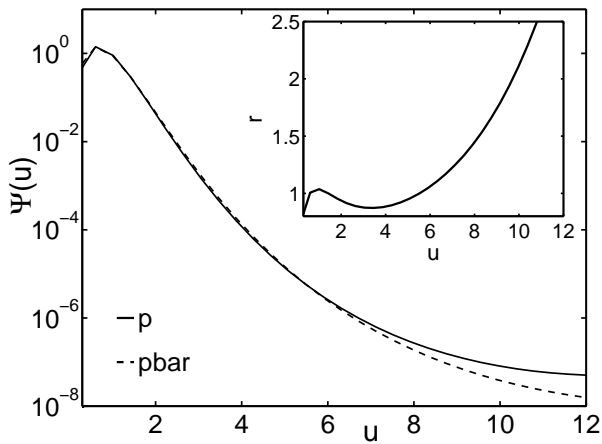}
\caption{Comparison between the scaling functions for protons and anti-protons
produced at mid-rapidity in Au+Au collisions at
RHIC with the scaling variable $u$. The inset if for the ratio
$\Psi_p(u)/\Psi_{\bar{p}}(u)$.}
\label{fig6}
\end{figure}

The scaling behaviors of the $p_T$ distribution functions for protons
and anti-protons can be tested experimentally from studying the ratio of
moments of the momentum distribution, $\langle p_T^n\rangle/\langle p_T\rangle^n
=\int_0^\infty u^n \Psi(u)udu$ for $n=2, 3, 4, \cdots$. From the determined normalized
 distributions, the ratio can be calculated by integrating over $u$ in  the range
 from 0 to 12, as mentioned above, and the results
are tabulated in TABLE II. The values of the ratio are independent of the parameters
$A$ and $K$ in the fitting process but only on the functional form
of the scaling distributions. If the scaling behaviors of particle distributions are
true, such ratios should be constants independent of the colliding centralities
and rapidities. For comparison, the corresponding values of the
ratio for pions produced in the same interactions, calculated in \cite{yang1},
are also given in TABLE II. Because of very small difference in the scaling
distributions for protons and anti-protons at small $u$, the ratio for protons
increases with $n$ at about the same rate as for anti-proton for small $n$. For large $n$, the
ratio for $p$ becomes larger than that for $\bar{p}$ because of the big difference
in the scaling functions for $p$ and $\bar{p}$ at large $u$. Because of the very strong
suppression of high transverse momentum proton production relative to that of pions,
the ratio for pions increases with $n$ much more rapidly
than for $p$ and $\bar{p}$.
\begin{table}
\def\tabcolsep{0.5cm}
\begin{tabular}{||c|c|c|c||}
\hline
 $n$ & $p$ & $\bar{p}$ & $\pi$\\ \hline
 2 & 1.194 & 1.215 & 1.65\\ \hline
 3 & 1.717 & 1.775 & 4.08\\ \hline
 4 & 2.978 & 3.064 & 14.4\\ \hline
 5 & 6.415 & 6.417 & 64.73\\ \hline
 6 & 19.045 & 17.253 & 373.82\\ \hline
\end{tabular}
\caption{Ratio of moments $\langle p_T^n\rangle/\langle p_T\rangle^n$ for
protons, anti-protons and pions produced in Au+Au collisions at RHIC.}
\end{table}

 Another important question is about the difference between the scaling functions
for protons in this paper and for pions in \cite{hy1,yang1}. Experiments at RHIC have shown
that the ratio of proton yield over that of pion increases with $p_T$ up to 1
in the region $p_T\leq 3$ GeV$/c$ and saturates in large $p_T$ region. This
behavior should be seen from the scaling functions for these two species of particles.
For the purpose of comparing the scaling distributions we define a ratio
\begin{equation}
R=\Psi_p(u)/\Psi_\pi(u)\ ,
\end{equation}
and plot the ratio $R$ as a function of $u$ in Fig. 7. The ratio increases with $u$,
when $u$ is small, reaches a maximum at $u$ about 1 and then decreases. Finally it decreases
slowly to about 0.1 for very large $u$. The highest value of $R$ is about 1.6, while
the experimentally observed $p$ over $\pi$ ratio is about 1 at $p_T\sim 3\ {\rm GeV}/c$.
The reason for this difference is two-fold. One is the normalization difference in
defining $R$ and the experimental ratio. Another lies in the different mean transverse
momenta $\langle p_T\rangle$'s for pions and protons with which the scaling variable $u$
is defined and used in getting the ratio $R$.

The existence of difference in the scaling distributions for different species of
particles produced in high energy collisions is not surprising, because the distributions
reflect the particle production dynamics which may be different for different particles.
In the quark recombination models \cite{hy,gr,fr} pions are formed by combining
a quark and an anti-quark while protons by three quarks. Because different
numbers of (anti)quarks participate in forming the particles, their scaling
distributions must be different. In this sense, our investigation results
urge more studies on particle production mechanisms.
 \begin{figure}[tbph]
\includegraphics[width=0.45\textwidth]{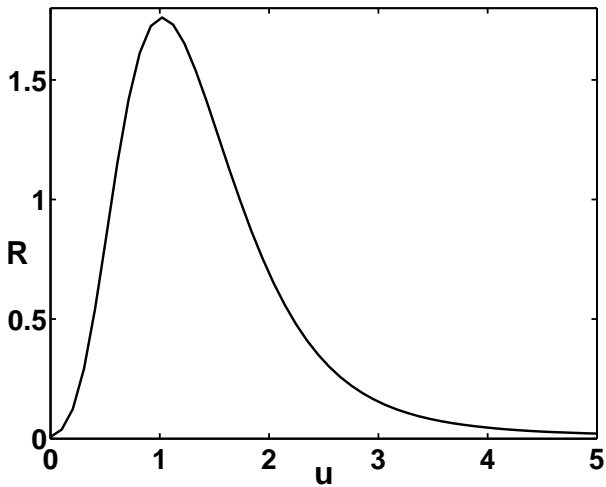}
\caption{Ratio $\Psi_p(u)/\Psi_\pi(u)$ between the scaling functions for protons and pions
produced in Au+Au collisions at RHIC as a function of the scaling variable $u$.
The pion scaling distribution is from \cite{hy1,yang1}.}
\label{fig7}
\end{figure}

\section{Discussions}
From above investigation we have found scaling distributions for protons and
anti-protons produced in Au+Au collisions at RHIC in both mid-rapidity and forward region.
The difference between those two scaling distributions is quite small, but they
differ a lot from that for pions and the ratio $\Psi_p/\Psi_\pi$
exhibits a nontrivial behavior.

Investigations in \cite{hy1,yang1} and in this paper have shown that particle
distributions can be put to the same curve by linear transformation on $p_T$.
Though we have not yet a uniform picture for the particle productions in high energy
nuclear collisions, the scaling behaviors can, in some sense, be compared to that from
the string fragmentation picture \cite{string}. In that picture if there are $n$ strings,
they may overlap in an area of $S_n$ and the average area for a string
is then $S_n/n$. It is shown that the momentum distributions can be related to
the case in $pp$ collisions also by a linear variable change $p_T\to
p_T ((S_n/n)_{\rm AuAu}/(S_n/n)_{pp})^{1/4}$. Viewed from that picture, our fitted
$K$ gives the degree of string overlap. The average area for a string in
most central Au+Au collisions is about 70 percent of that in peripheral
ones from the values of $K$ obtained from fitting the spectra of proton. If string
fragmentation is really the production mechanism for all species
of particles in the collisions, one can expect that the overlap degree obtained
is the same from the changes of spectrum of any particle. In the language
in this work, values of $K$ are expected the same for pions, protons and other
particles in the string fragmentation picture for particle production.
Our results show the opposite. Comparing the values of $K$ from \cite{hy1}
and this work, one can see that for pion spectrum $K$ is larger for more peripheral
collisions  but smaller for proton and anti-proton spectra.  Our results indicate that
other particle production mechanisms may also provide ways to the scaling distributions.
Obviously more detailed studies, both theoretically and experimentally, are needed.

\acknowledgments
This work was supported in part by the National Natural
Science Foundation of China under Grant Nos. 10635020 and 10475032, by
the Ministry of Education of China under Grant No. 306022 and project IRT0624.

\end{document}